\documentclass[3p,times,procedia]{elsarticle}
\usepackage{nupha_ecrc}

\volume{00}

\firstpage{1}

\journalname{Nuclear Physics A}

\runauth{}

\jid{nupha}

\jnltitlelogo{Nuclear Physics A}

\usepackage{amssymb}
\usepackage{floatrow}

\usepackage{lineno}
\usepackage{caption}
\usepackage{subcaption}

\usepackage[figuresright]{rotating}

\newcommand{\figref}[1]{Fig.~\ref{#1}}
\newcommand{\Figref}[1]{Figure~\ref{#1}}
\newcommand{\dEdx}{d$E$/d$x$}

\begin{document}

\begin{frontmatter}



\dochead{XXVIIIth International Conference on Ultrarelativistic Nucleus-Nucleus Collisions\\ (Quark Matter 2019)}

\title{Recent results on net-baryon fluctuations in ALICE}

 \author[label1]{Mesut Arslandok (on behalf of the ALICE collaboration)}
 \address[label1]{Physikalisches Institut, Universit\"at Heidelberg, Heidelberg, Germany}

\author{}

\address{}

\begin{abstract}
Recent results on the analysis of event-by-event net-baryon number fluctuations in Pb--Pb collisions at $\sqrt{s_{\mathrm{NN}}} = 2.76$ and $5.02$~TeV are presented. The cumulants of the net-proton distributions, proxies for the net-baryon distributions, up to third order are discussed. The experimental results are compared with HIJING and EPOS model calculations and the dependence of fluctuation measurements on the phase-space coverage is addressed in the context of calculations from Lattice QCD (LQCD) and the Hadron Resonance Gas (HRG) model.
\end{abstract}


\end{frontmatter}


\section{Introduction} \label{sec:Introduction}
\begin{figure}
  \floatbox[{\capbeside\thisfloatsetup{capbesideposition={right,top},capbesidewidth=4cm}}]{figure}[\FBwidth]
  {\caption{Pseudorapidity dependence of the second cumulants of net-pions (left) and net-kaons (right) normalized to the corresponding second cumulants of the Skellam distributions. The ALICE data are shown by black markers while the blue solid and dashed lines indicate the HIJING~\cite{Gyulassy:1994ew} model calculations with and without resonance contributions, respectively.}\label{fig:2nd_cumulants_Resonance_Effect}}
  {\includegraphics[width=10cm]{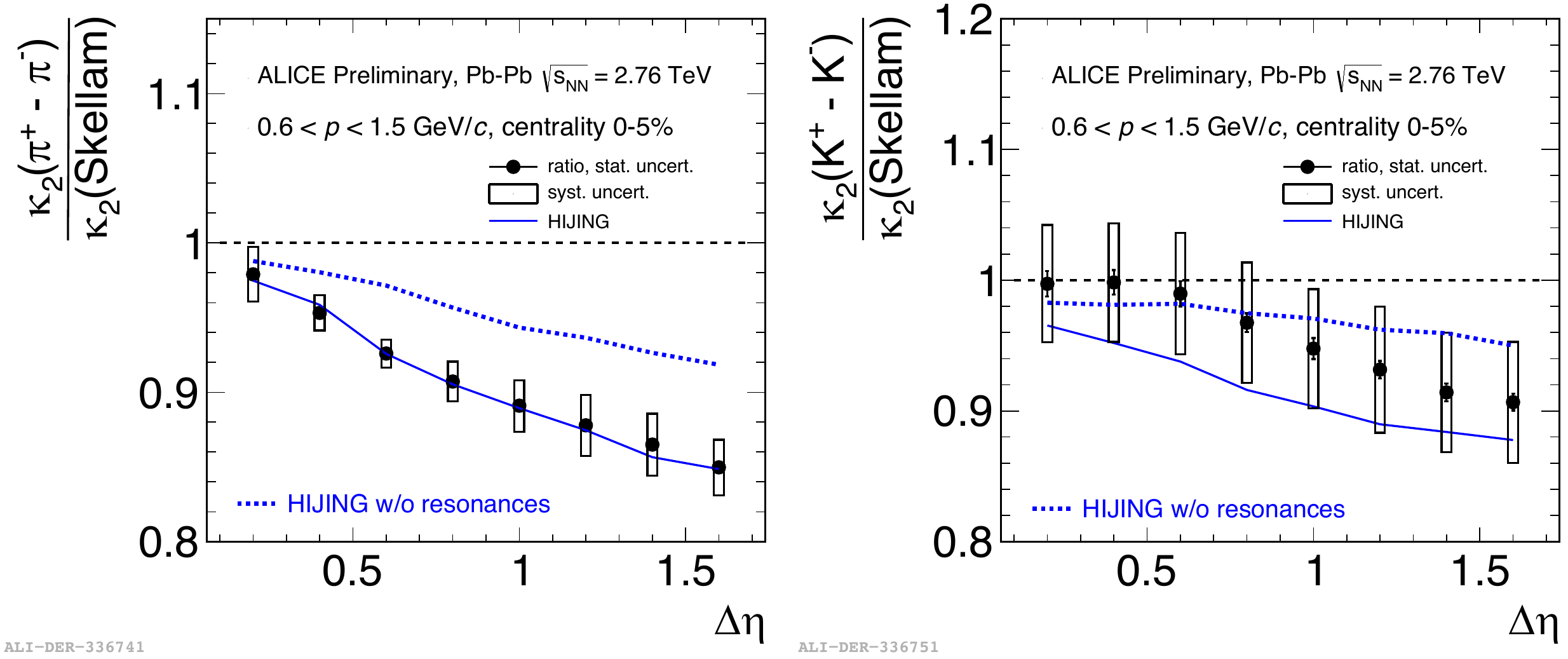}}
\end{figure}

One of the key goals of nuclear collision experiments is to map the phase diagram of strongly interacting matter. At LHC energies there would be, for vanishing light quark masses, a temperature-driven genuine phase transition of second order between the hadron gas and the quark--gluon plasma. For realistic quark masses, however, this transition becomes a smooth cross over. Nevertheless, due to the small masses of current quarks one can still probe critical phenomena at LHC energies, which can be confronted with ab-initio LQCD calculations at vanishing baryon chemical potential. Indeed, recent LQCD calculations \cite{LQCD1,Bazavov:2018mes,LQCD2} exhibit a rather strong signal for the existence of a pseudo-critical chiral temperature ($T_{\mathrm{pc}}^{\mathrm{LQCD}}$) of about 156~MeV, which is consistent with the chemical freeze-out  temperature ($T_{\mathrm{fo}}^{\mathrm{ALICE}}$) extracted by the analysis of hadron multiplicities~\cite{HRG} measured by the ALICE experiment. 
For a thermal system in a fixed volume, within the Grand Canonical Ensemble (GCE) formulation, the event-by-event fluctuations of conserved charges are related to thermodynamic susceptibilities, which are calculable in the LQCD framework. The susceptibilities are defined as the partial derivatives of the reduced pressure with respect to the reduced chemical potential, $\chi_{n}^{N=\mathrm{B,S,Q}}=\partial^{n} (P/T^{4})/\partial (\mu_{N}/T)^{n}$, 
where B, S and Q correspond to baryon number, strangeness and electric charge. In order to get rid of volume and temperature terms, which enter into this equation, the charge susceptibilities are studied experimentally in terms of the ratios of cumulants~\cite{Gavai:2010zn}.
\\
\indent 
Measurements of net-proton, net-kaon and net-pion cumulants are used to study net-baryon number, net-strangeness and net-electric charge fluctuations, respectively. Since net-kaon and net-pion multiplicities are dominated by resonance decays (see difference in HIJING model calculations with and without resonances in \figref{fig:2nd_cumulants_Resonance_Effect}), e.g., $\phi \rightarrow K^{+}K^{-}$ and $\rho \rightarrow \pi^{+}\pi^{-}$, the analysis of net-electric charge and net-strangeness remains a challenge for both experimentalists and theorists. On the other hand, there are no resonances which decay into $p\overline{p}$ with a sizable branching ratio, therefore net-proton measurements are the best candidates to study charge susceptibilities~\cite{Kitazawa:2012at}.
%
%
%
\section{The experimental data and analysis method} \label{sec:dataAndMethod}
The analysis presented in this report is based on about 13 and 78 million minimum-bias Pb–Pb collisions at $\sqrt{s_{\mathrm{NN}}}=2.76$ and $5.02$~TeV collected in the years 2010 and 2015, respectively. The detectors used are the Time Projection Chamber (TPC) for tracking and particle identification and the Inner Tracking System (ITS) for tracking and precise vertex determination. The definition of the collision centrality is based on the charged-particle multiplicity measured in two forward scintillator hodoscopes (V0 detectors) that are located on either side of the interaction point~\cite{Abelev:2013qoq}. The particle identification is obtained via the specific energy loss (\dEdx) measured in the TPC. Since the kinematic acceptance plays an important role in the fluctuation of conserved charges, the analysis is performed differentially as a function of the pseudorapidity acceptance ($\Delta \eta$) and within two different momentum ranges: $0.6<p<1.5$~GeV/\textit{c} and $0.6<p<2$~GeV/\textit{c}. 
\\
\indent The net-proton cumulants are reconstructed using a novel approach, the Identity Method~\cite{Rustamov:2012bx,Arslandok:2018pcu}, which overcomes incomplete particle identification caused by the overlapping \dEdx\ distributions. Hence, requiring additional detector information (e.g. Time-Of-Flight (TOF)), which substantially reduces the particle detection efficiencies, is avoided, and thus the detection efficiencies are kept as high as possible. One should note that small efficiencies not only reduce the dynamical fluctuations of interest but also increase experimental uncertainties arising from the efficiency correction procedure~\cite{Luo:2018ofd}. The detection efficiency of protons within the kinematic acceptance used in this analysis is about $83\%$ and almost independent of the collision centrality, momentum and pseudorapidity. The remaining efficiency loss is corrected under the assumption of binomial track loss~\cite{Nonaka:2017kko}, which has been extensively tested using simulated Monte Carlo events passed through a GEANT model of the ALICE detector. Further details on the application of the Identity Method on the ALICE data can be found in \cite{Acharya:2017cpf}.
\section{Results} \label{sec:Results}
\begin{figure}[htbp]
  \centering
  \begin{subfigure}{.5\textwidth}
    \centering
    \includegraphics[width=7.5cm]{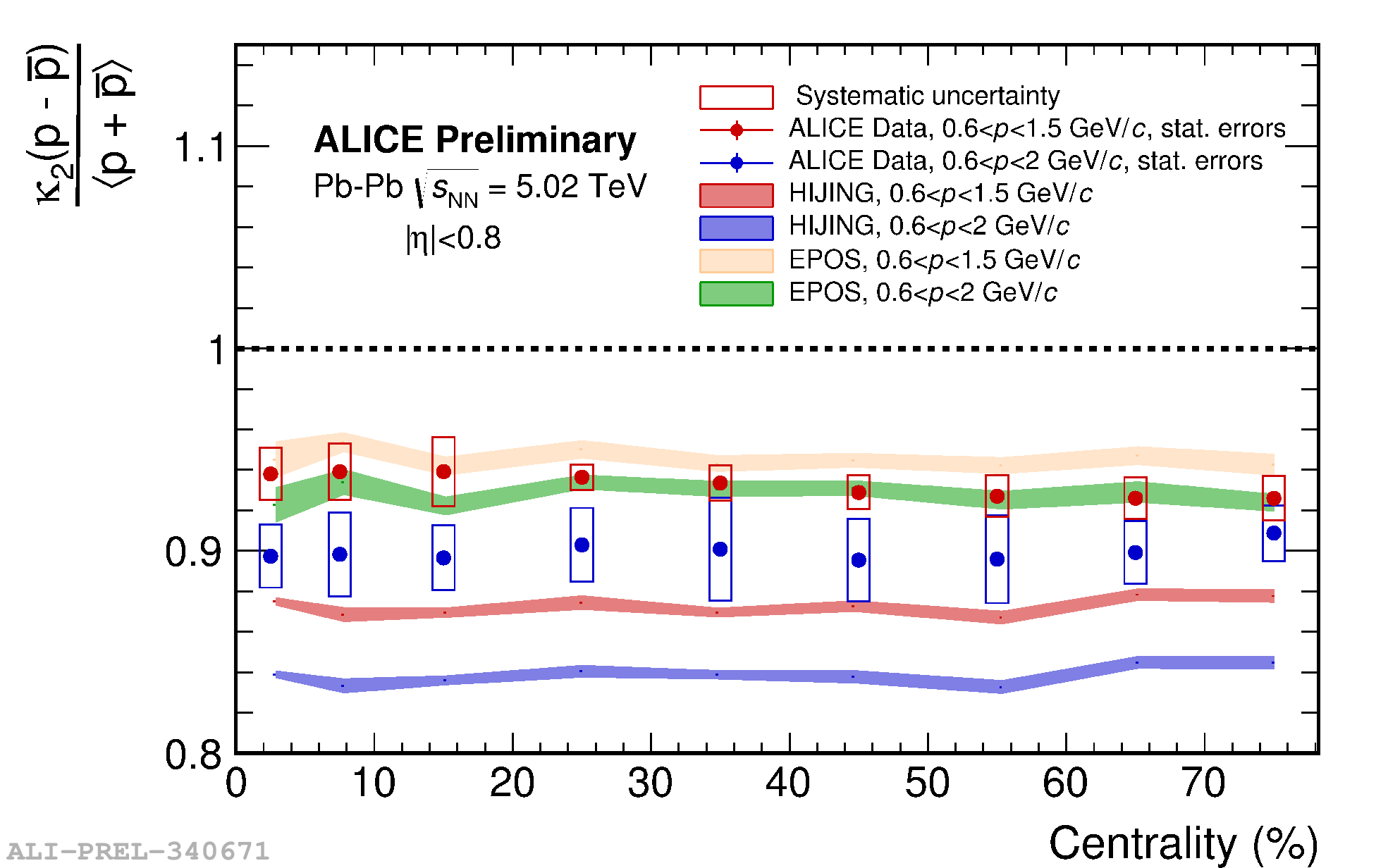}
  \end{subfigure}%
  \begin{subfigure}{.5\textwidth}
    \centering
    \includegraphics[width=7.5cm]{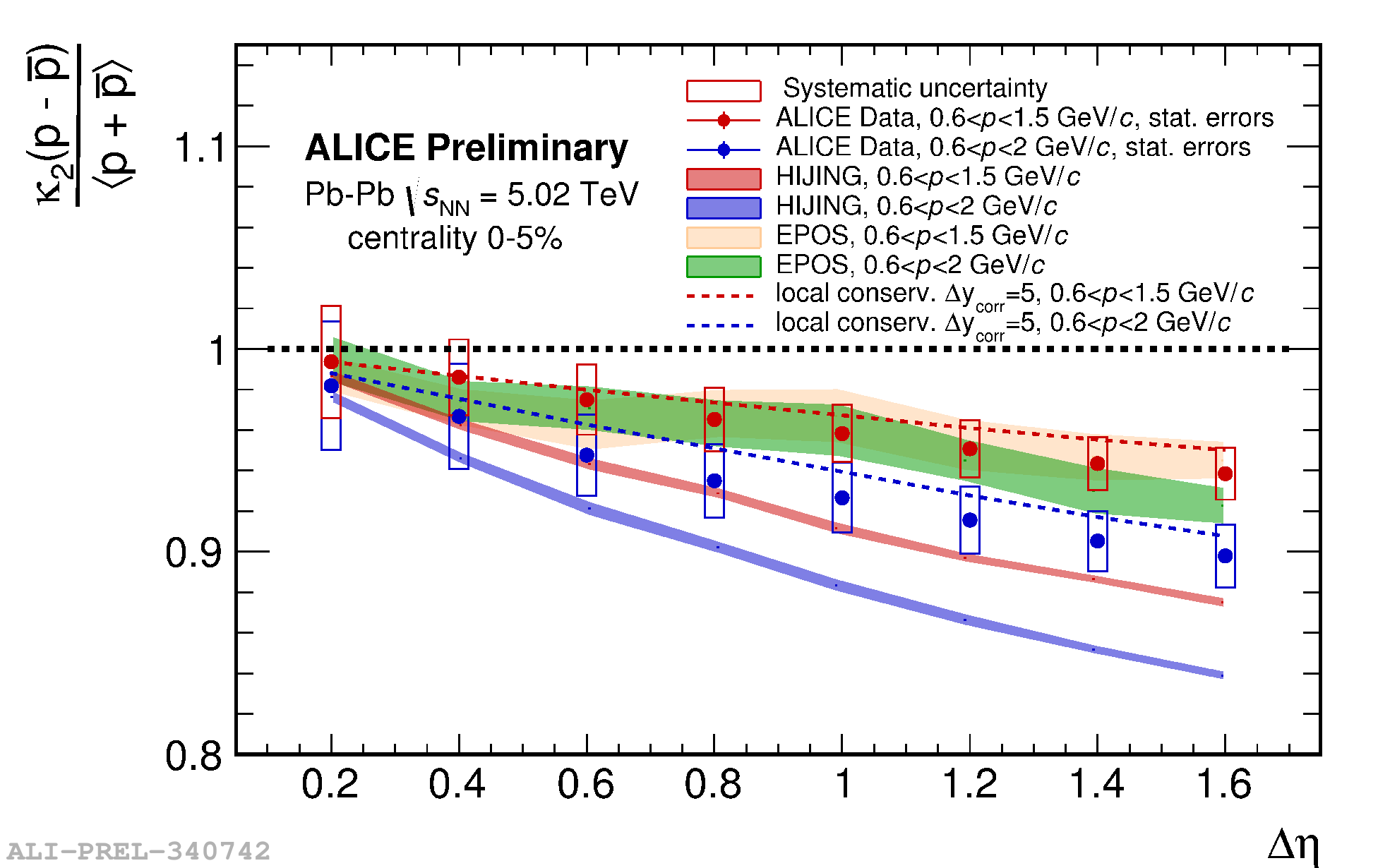}
  \end{subfigure}
  \caption{(Color online) Centrality (left) and pseudorapidity (right) dependence of the normalized second cumulants of net-protons. The ALICE data are shown by red and blue markers for two different kinematic acceptances in momentum space; $0.6<p<1.5$~GeV/\textit{c} and $0.6<p<2$~GeV/\textit{c}, respectively. The colored shaded areas indicate the HIJING~\cite{Gyulassy:1994ew} and EPOS~\cite{Pierog:2013ria} model calculations and the dashed lines represent the predictions from the model with local baryon number conservation~\cite{Braun-Munzinger:2019yxj}.}
  \label{fig:2nd_Cumulants_MomentumRange_dep}
\end{figure}

\begin{figure}[htbp]
   \centering
  \begin{subfigure}{.5\textwidth}
    \centering
    \includegraphics[width=7cm]{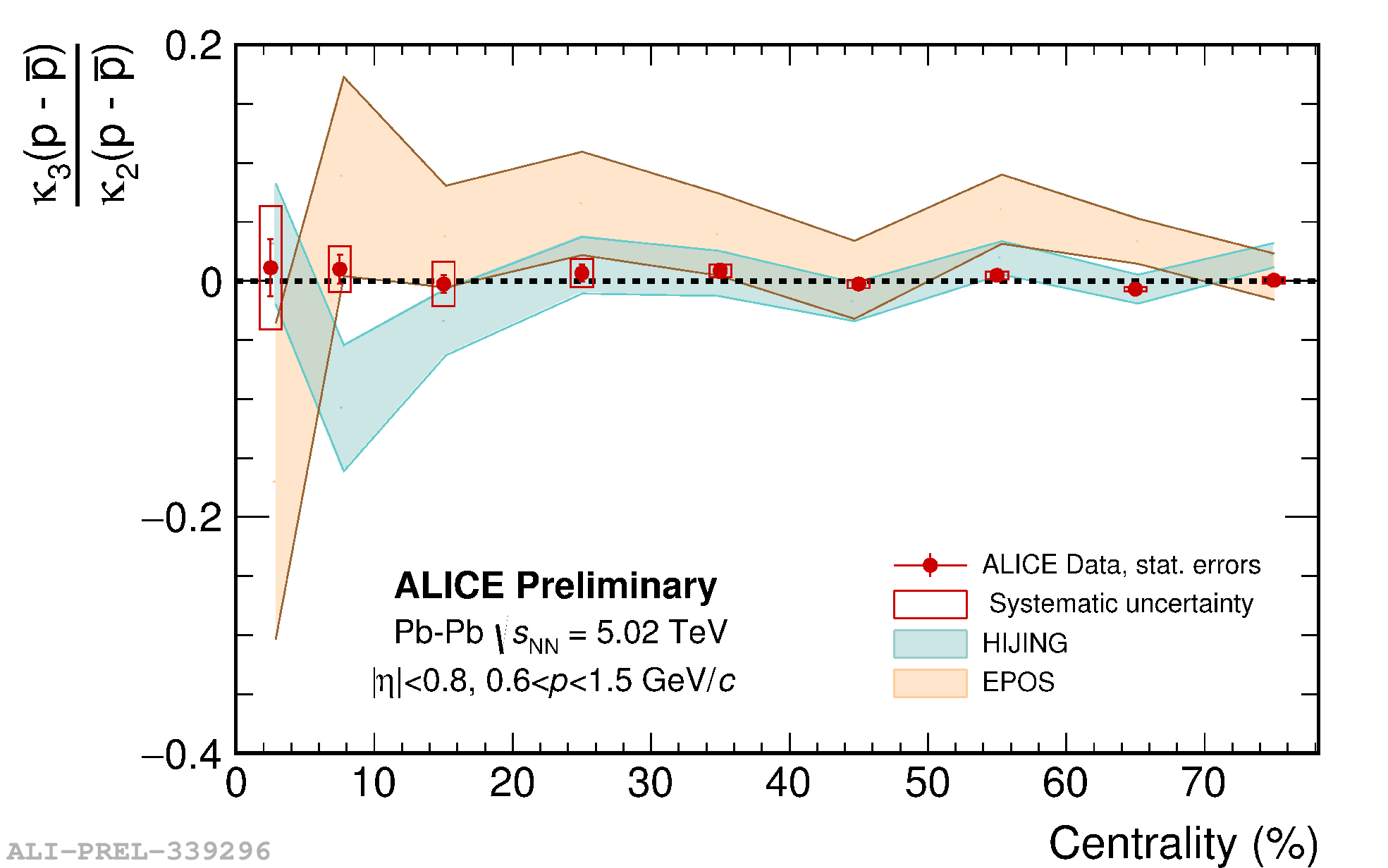}
  \end{subfigure}%
  \begin{subfigure}{.5\textwidth}
    \centering
    \includegraphics[width=7cm]{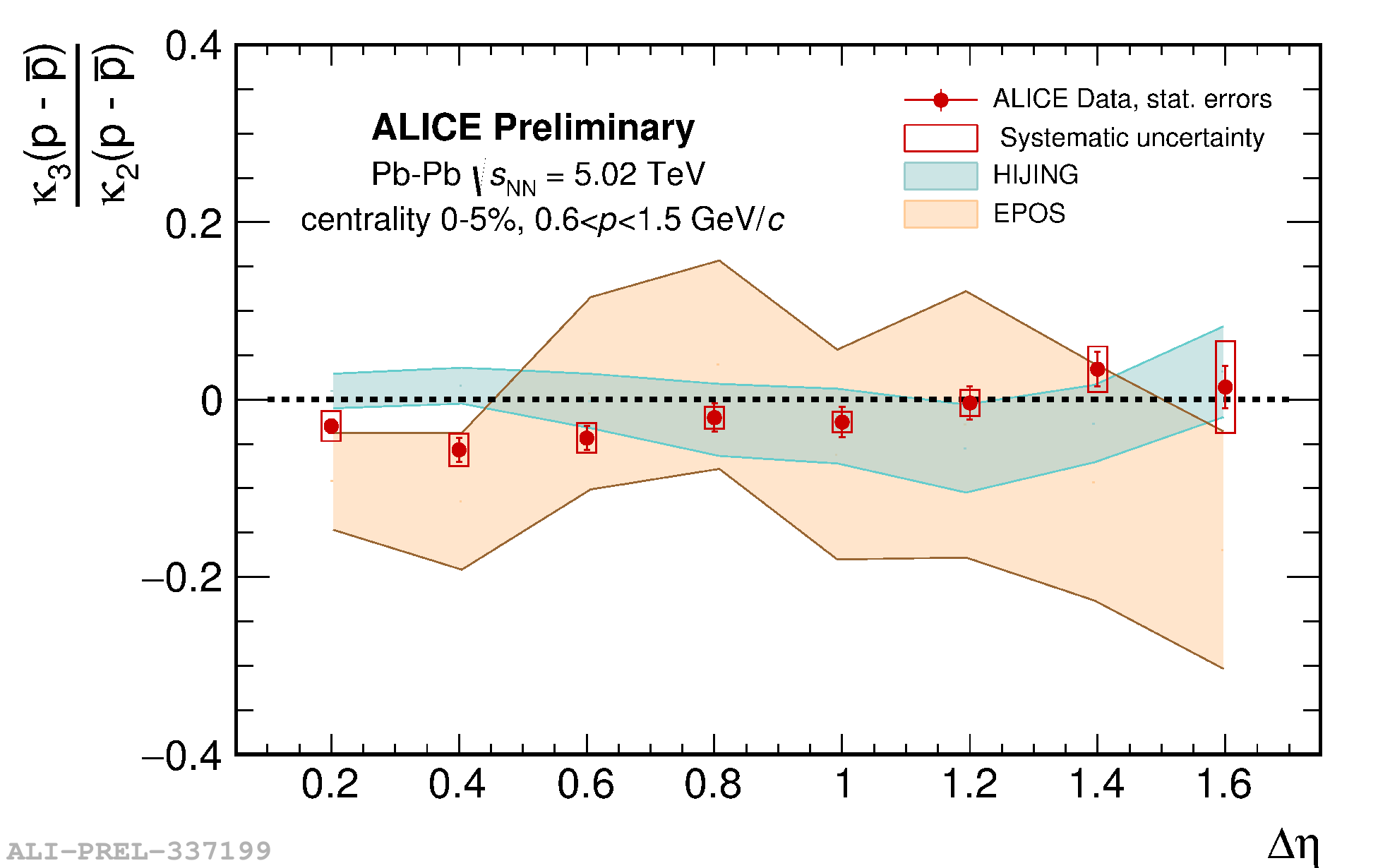}
  \end{subfigure}
  \caption{(Color online) Centrality (left) and pseudorapidity (right) dependence of the ratio of third to second cumulants for net-protons at $\sqrt{s_{\mathrm{NN}}}=5.02$~TeV. The ALICE data are shown by red markers, while the colored shaded areas indicate the HIJING~\cite{Gyulassy:1994ew} and EPOS~\cite{Pierog:2013ria} model calculations.}
  \label{fig:3rd_Cumulants}
\end{figure}

At the temperature $T_{\mathrm{pc}}^{\mathrm{LQCD}}=156~\mathrm{MeV}\approx T_{\mathrm{fo}}^{\mathrm{ALICE}}$ both HRG and LQCD predict a Skellam behavior for the second and third net-baryon cumulants, while the fourth cumulants from LQCD are significantly below the corresponding Skellam baseline~\cite{Karsch, Kaczmarek:2017hfx}. The Skellam distribution is defined as the probability distribution of the difference of two random variables, each generated from statistically independent Poisson distributions. For net-protons, the $n$th cumulants of the Skellam distribution are given by; $\kappa^{\mathrm{Skellam}}_{n}(\mathrm{p}-\mathrm{\overline{p}}) = \langle \mathrm{p} \rangle + (-1)^{n} \langle \mathrm{\overline{p}} \rangle$, where $\langle \mathrm{p} \rangle$ and $\langle \mathrm{\overline{p}} \rangle$ are the mean values of the proton and anti-proton multiplicity distributions, respectively. Here one should note that conserved quantities fluctuate only within a limited kinematic acceptance and are analyzed within the GCE framework, where net-baryon number is conserved only on average. However, in the limit of very small kinematic acceptances, dynamical correlations are suppressed and fluctuations asymptotically approach the Skellam limit. At LHC energies, the numbers of protons and anti-protons produced at mid-rapidity are equal~\cite{Abelev:2013vea}, and thus the normalized cumulants of the Skellam distribution with respect to its second cumulants are 0 for odd and 1 for even cumulants. 
\begin{figure}[ht]
  \centering
  \begin{subfigure}{.5\textwidth}
    \centering
    \includegraphics[width=7.5cm]{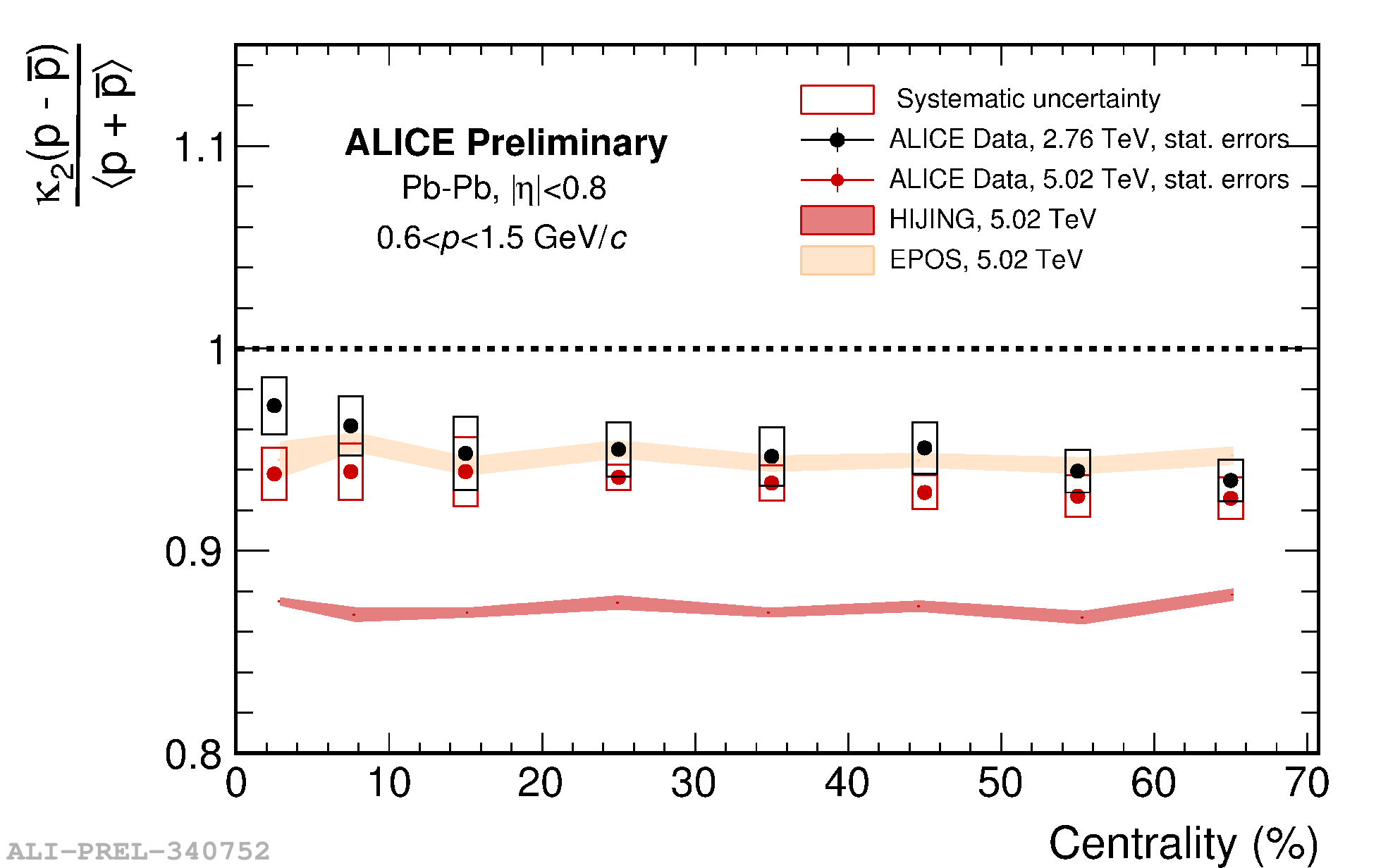}
  \end{subfigure}%
  \begin{subfigure}{.5\textwidth}
    \centering
    \includegraphics[width=7.5cm]{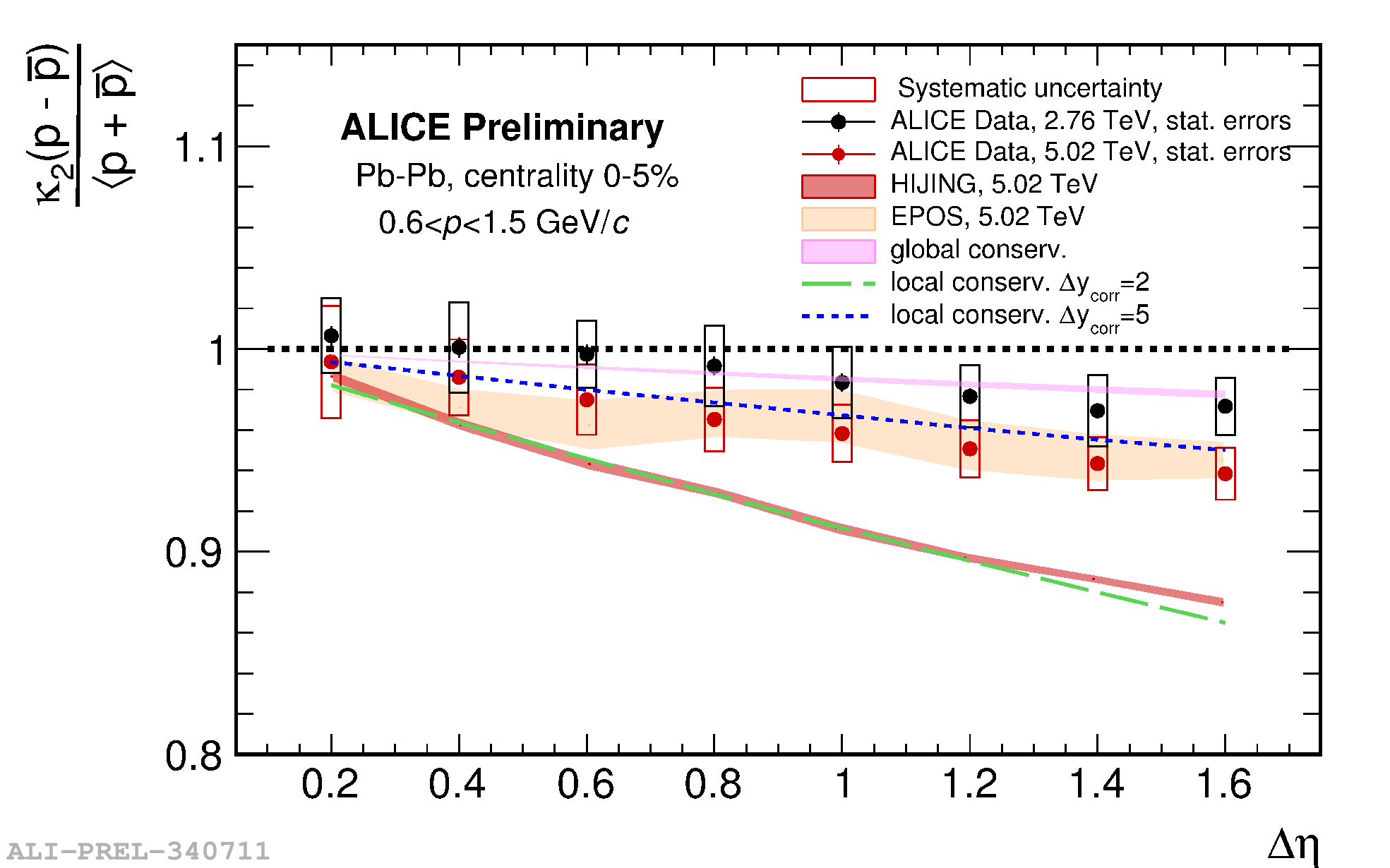}
  \end{subfigure}
  \caption{(Color online) Centrality (left) and pseudorapidity (right) dependence of the normalized second cumulants of net-protons. The ALICE data are shown by black and red markers for $\sqrt{s_{\mathrm{NN}}}=2.76$ and $5.02$~TeV, respectively, while the colored shaded areas indicate the HIJING~\cite{Gyulassy:1994ew} and EPOS~\cite{Pierog:2013ria} model calculations at $\sqrt{s_{\mathrm{NN}}}=5.02$~TeV. On the right panel global  baryon number conservation is depicted as the pink band and the dashed lines represent the predictions from the model with local baryon number conservation~\cite{Braun-Munzinger:2019yxj}.}
  \label{fig:2nd_Cumulants_Energy_dep}
\end{figure}
\\
\indent 
In \figref{fig:2nd_Cumulants_Energy_dep} and \figref{fig:2nd_Cumulants_MomentumRange_dep} the measured centrality and $\Delta \eta$ dependence of the net-proton second cumulants normalized to the corresponding second cumulants of the Skellam distribution are shown. The results for the two different energies, which are independent of centrality, agree within the experimental uncertainties and show a deviation from the Skellam baseline (\figref{fig:2nd_Cumulants_Energy_dep} left panel). The amount of the deviation is in good agreement with a model assuming baryon number conservation~\cite{Braun-Munzinger:2019yxj,Braun-Munzinger:2016yjz,Acharya:2019izy}, where the data suggest long range rapidity correlations ($\Delta y_{corr}>5$) between protons and anti-protons, therefore arising from the early phase of the collision~\cite{Dumitru:2008wn} (\figref{fig:2nd_Cumulants_Energy_dep} right panel). The EPOS~\cite{Pierog:2013ria} model calculations agree with the data while HIJING favors a smaller correlation length ($\Delta y_{corr}=2$) presumably due to string breaking, implemented using the Lund string model. Both data and the model calculations for two different kinematic acceptances in $p$ are independent of centrality (\figref{fig:2nd_Cumulants_MomentumRange_dep} left panel), while they show an increasing deviation from the Skellam baseline with increasing kinematic acceptance in both $\eta$ and $p$ (right panels of \figref{fig:2nd_Cumulants_Energy_dep} and \figref{fig:2nd_Cumulants_MomentumRange_dep}). This is also in agreement with the predictions in \cite{Braun-Munzinger:2019yxj}.
\\
\indent 
\Figref{fig:3rd_Cumulants} shows the ratio of the measured third cumulants of net-protons to the second cumulants. After correction for efficiencies they are zero within uncertainties, in line with expectations from the HRG and LQCD. In calculations using the HIJING and EPOS models the numbers of protons and anti-protons within the current experimental acceptance are equal. Therefore the resulting third order cumulants are zero within uncertainties for all centralities and $\Delta \eta$. The experimentally achieved precision of better than 5\% is promising for the analysis of higher cumulants in the near future. A factor of 10 more data was collected in 2018, making accessible the fourth cumulants, and a factor of 100 more is expected from 2021 onwards.
%
\vspace{-0.5cm}
\section{Conclusions} \label{sec:Conclusions}
It is shown that net-pion and net-kaon fluctuations are strongly dominated by resonance contributions, which is not the case for protons, making them a good measure for baryon number fluctuations. The second and third cumulants of the net-proton distribution are presented as function of centrality and kinematic acceptance for two different energies, where the Identity Method was applied to obtain the third cumulants for the first time. The net-proton second cumulants are in agreement with the corresponding second cumulants of the Skellam distribution after accounting for baryon number conservation. In this context, the ALICE data suggest long range rapidity correlations ($\Delta y_{corr}>5$) between protons and anti-protons that arise from the early phase of the collision. The third cumulants are consistent with 0 within the experimental uncertainties. The ALICE data are found to be in agreement with the LQCD expectations up to the third order cumulants. The fourth and higher order cumulants, where net-baryons from LQCD are significantly below the corresponding Skellam baseline, are currently being studied. 
\vspace{-0.3cm}
\section*{Acknowledgements}
This work is part of and supported by the DFG Collaborative Research Centre ``SFB 1225 (ISOQUANT)''.
\vspace{-0.8cm}
%
%
\bibliographystyle{elsarticle-num}
\bibliography{biblio.bib}
\nopagebreak
%






\end{document}